\documentclass[citeautoscript,floatfix,aps,prb,twocolumn,superscriptaddress]{revtex4-1}
\usepackage{graphicx}
\usepackage{amsmath}
\usepackage{amssymb}
\usepackage{dcolumn}
\usepackage[version=4]{mhchem}
\usepackage{latexsym}
\usepackage{rotating}
\usepackage{epstopdf}
\usepackage{siunitx}
\usepackage[usenames,dvipsnames]{xcolor}
\usepackage{float}
\usepackage{soul}
\usepackage{epsfig}
\usepackage{psfrag}
\usepackage{natbib}
\usepackage{bm}
\usepackage{eucal}
\usepackage{mathrsfs}
\usepackage{braket}
\usepackage{enumerate}
\usepackage{longtable}
\usepackage{bm}
\usepackage{hyperref}
\usepackage{amsfonts}
\usepackage{dcolumn}
\usepackage{bm}
\usepackage{changes}

\newcommand{\be}{\begin{equation}}
	\newcommand{\ee}{\end{equation}}
\newcommand{\bn}{\begin{eqnarray}}
	\newcommand{\en}{\end{eqnarray}}
\newcommand{\qsgwl}{QS$G\widehat{W}$}

\def\x2y2{{x^2-y^2}}

\usepackage{color} 

\usepackage{hyperref}
\hypersetup{
	colorlinks=true,final=true,
	linkcolor=red,
	citecolor=blue,
	filecolor=blue,
	urlcolor=blue,
}

\begin{document}
	
	\title{One-particle and excitonic band structure in cubic Boron Arsenide}
	\author{Swagata Acharya}
	\affiliation{National Renewable Energy Laboratories, Golden, CO 80401, USA}
	\affiliation{Institute for Molecules and Materials, Radboud University, NL-6525 AJ Nijmegen, The Netherlands}
	\email{swagata.acharya@nrel.gov}
	\author{Dimitar Pashov}
	\affiliation{ King's College London, Theory and Simulation of Condensed Matter,
		The Strand, WC2R 2LS London, UK}
	\author{Mikhail I Katsnelson}
	\affiliation{Institute for Molecules and Materials, Radboud University, NL-6525 AJ Nijmegen, The Netherlands}
	\author{Mark van Schilfgaarde}
	\affiliation{National Renewable Energy Laboratories, Golden, CO 80401, USA}
	\affiliation{ King's College London, Theory and Simulation of Condensed Matter,
		The Strand, WC2R 2LS London, UK}
	
	\begin{abstract}
Cubic BAs has received recent attention for its large electron and hole mobilities and large thermal conductivity. This
is a rare and much desired combination in semiconductor industry: commercial semiconductors typically have high electron
mobilities, or hole mobilities, or large thermal conductivities, but not all of them together.
Here we report predictions from an advanced self-consistent many body perturbative theory and show that with respect to
one-particle properties, BAs is strikingly similar to Si.  There are some important differences, notably there is an unusually small variation in the valence band masses .  With respect to two-particle properties,
significant differences with Si appear. We report the excitonic spectrum for both \textbf{q}=0 and finite \textbf{q},
and show that while the direct gap in cubic BAs is about 4\,eV, dark excitons can be observed down to about
$\sim$1.5\,eV, which may play a crucial role in application of BAs in optoelectronics.
\end{abstract}

\maketitle

\section{Introduction}
Zincblende BAs has received considerable attention recently~\cite{shin2022} because
of its large ambipolar mobility and large thermal conductivities.  It is a rare yet much desired combination.  Silicon
--- the foundation for much of modern day technology --- has poor hole mobility and also is a bad thermal conductor and,
hence, suffers from overheating issues.  BAs has thermal conductivity that is nearly ten times higher than silicon and
nearly half of that of diamond.  A series of theoretical~\cite{theory2} and experimental works~\cite{tc1,tc2,tc3} almost
unambiguously established these facts.  However,  experimental determination of the ambipolar mobilities
have been more difficult to establish.  A density-functional calculation in
2018~\cite{theory1} predicted high electron and hole mobilities of $\mu_{e}$=1400 cm$^{2}$\,V$^{-1}$\,s$^{-1}$ and
$\mu_{h}$=2100\,cm$^{2}$\,V$^{-1}$\,s$^{-1}$ .  However, a subsequent experimental work found the hole
mobility to be as low as 22 cm$^{2}$\,V$^{-1}$\,s$^{-1}$.  Defects were thought to be the key factor limiting the hole
mobility and it demanded an experimental approach where thermal and electrical conductivities could be measured for the
intrinsic sample without being defect-limited.  Shin et al.~\cite{shin2022}  achieved this goal by employing
transient grating method that  simultaneously measured  thermal ($\kappa_{RT}$) and electrical
conductivities (or, equivalently, $\mu_{e},\mu_{h}$) at different spots (and a small region around them) of the sample
which are free from defects and provide the information of the intrinsic mobilities.  With that they
found $\mu_{h}{>}1000$\,cm$^{2}$\,V$^{-1}$\,s$^{-1}$, ambipolar mobility ${2\mu_{e}\mu_{h}}/({\mu_{e}+\mu_{h}})$ of
1600\,cm$^{2}$\,V$^{-1}$\,s$^{-1}$ and $\kappa_{RT}$ of 1200 Wm$^{-1}$K$^{-1}$. Recent works by Yue et al.~\cite{ref11} and Choudhry et al.~\cite{ref12} also establish these facts from transient reflectivity microscopy and ultrafast electron microscopy, respectively.  These observations make cubic BAs an
attractive candidate for next generation semiconductor based technologies.

\section{Results and Discussion}

\subsection{Electronic properties}

\subsubsection{Electronic band gap}However, still there exists some ambiguities related to its electronic band gap, the anisotropies of the hole and
electron mass tensors, dielectric constant and excitonic properties. Several experimental and theoretical works have
reported significantly different numbers for these properties (see Table 1 in the work by Buckeridge and
Scanlon~\cite{buckeridge}).  The calculations from Buckeridge and Scanlon are performed on a zinc-blende structure of
BAs. Buckeridge and Scanlon reports an indirect band gap of 1.67 eV from their calculations. Older experimental studies
report band gap of $\sim$1.5 eV \cite{old1,old2,old3}, though a more recent measurement reports a gap of
1.82\,eV~\cite{Kang19}, and another 2.02\,eV from high-quality millimeter-sized crystals.~\cite{song2020}.  A single
shot $G_{0}W_{0}$ calculation performed with BerkleyGW yielded a gap of 2.07\,eV and another work reports a gap of 1.78 eV~\cite{ref21}.  Here we use quasiparticle
self-consistent $GW$ theory (QS$GW$) \cite{qsgw,pashov2020questaal}, which, in contrast to conventional $GW$ methods, modifies the charge
density and is determined by a variational principle \cite{variational}, and \qsgwl\ \cite{Cunningham2023} in which the
screened coulomb interaction $W$ is computed including vertex corrections (ladder diagrams) by solving a Bethe–Salpeter
equation (BSE) within Tamm-Dancoff approximation \cite{hirata1999}. Crucially, QS\emph{GW} and \qsgwl\ are fully
self-consistent in both self-energy $\Sigma$ and the charge density \cite{acharya2021importance}. $G$, $\Sigma$, and
$\widehat{W}$ are updated iteratively until all of them converge.  Zincblende semiconductors are described with
uniformly high fidelity in \qsgwl~\cite{Cunningham2023}, and we can expect similar agreement for BAs. We find the
electronic indirect band gap to be 2.00 eV within QS\emph{GW} and 1.85 eV within \qsgwl, indicating that vertex
corrections to the \emph{GW} self-energy are modest, as is typically found in III-V semiconductors.  (Since
\qsgwl\ systematically underestimates the $\Gamma$-X transition by 0.1 to 0.15\,eV in zincblende
semiconductors~\cite{Cunningham2023} the true bandgap is probably closer to 2.0\,eV).

\begin{table}[!ht]
	\centering
	\caption{Conduction band properties of BAs, at different levels of the theory:
		bandgaps, effective masses and nonparabolicity parameter $\alpha$.
		Si results are shown for comparison.  Experimental masses are taken from
		Ref.~\onlinecite{Hensel65}; the direct gap from Ref.~\onlinecite{Lautenschlager87}.}
	\label{tab:electron}
	\begin{tabular}{|@{\hspace{0.5em}}l@{\hspace{0.3em}}|@{\hspace{0.3em}}cc@{\hspace{0.3em}}|@{\hspace{0.3em}}cc@{\hspace{0.3em}}|@{\hspace{0.3em}}c@{\hspace{0.3em}}|}
		\toprule
		Theory         & $E_{G}$ & $E_{G}$(dir) & $m_{||}$ & $m_{\perp}$  & $\alpha_{\perp}$  \\
		\hline
		LDA            &  1.07   &  3.04       &   1.23   &   0.26    &  $-7.1$  \\
		QS\emph{GW}    &  2.00   &  4.26       &   0.96   &   0.25    &  $-7.0$   \\
		\qsgwl         &  1.85   &  4.08       &   1.03   &   0.25    &  $-7.4$   \\
		\qsgwl, Si     &  1.22   &  3.23       &   0.92   &   0.19    &  \hbox{\hskip 8pt}0.2 \\
		Expt, Si       &  1.17   &  3.35       &   0.91   &   0.19    &  \\
		\hline
	\end{tabular}
\end{table}

\begin{table}[h]
	\caption{\qsgwl\ Luttinger parameters for the valence band of BAs, compared to similar zincblende semiconductors.
	}
	\noindent
	\begin{tabular}{|r@{\hspace{0.1em}}|@{\hspace{0.2em}}c|@{\hspace{0.1em}}ccc|ccc|}
		\hline
		\vbox{\vskip 2pt}
		& & \multicolumn{3}{c|}{\qsgwl}
		&\multicolumn{3}{c|}{Expt}\\
		& $E_{G}$ &$\gamma_1$ &$\gamma_2$ &$\gamma_3$ &$\gamma_1$ &$\gamma_2$  &$\gamma_3$\\
		\hline
		\vbox{\vskip 12pt}
		Si\footnote{compilation in Ref.~\onlinecite{LandoltBornsteinVol41A1b}}
		& 1.13    & 4.24      &  0.32     &  1.42     & 4.26-4.29 &  0.34-0.38 &  1.45-1.56 \\
		AlAs\footnote{compilation in Ref.~\onlinecite{Vurgaftman01}}
		& 2.14    & 3.99      &  0.89     &  1.45     & 3.42–3.44 &  0.67–1.23 &  1.17–1.57 \\
		GaAs$^\mathrm{b}$
		& 1.63    & 6.75      &  1.83     &  2.74     & 6.79–7.20 &  1.90–2.88 &  2.68–3.05 \\
		BAs{\hskip 6pt}
		& 1.85    & 4.33      &  0.05     &  1.29     & & & \\
		\hline
	\end{tabular}
	\label{tab:luttinger}
\end{table}

\begin{figure*}[htb]
	\includegraphics[scale=0.65]{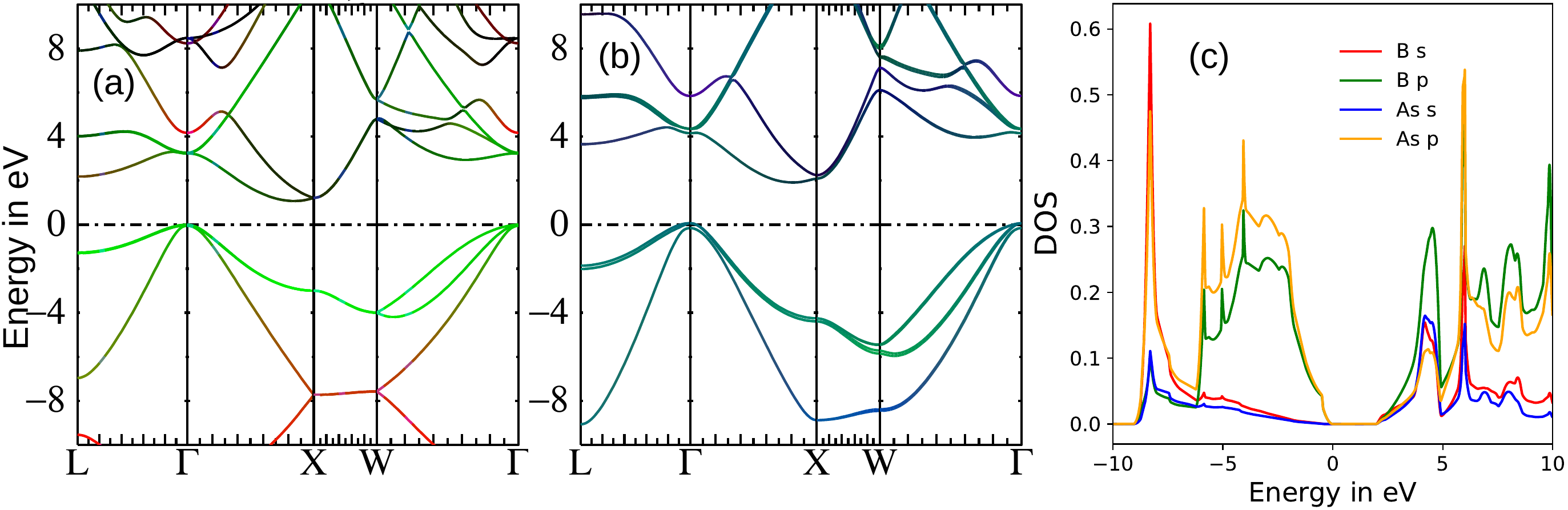}%
	\caption{{\bf Energy bands:} \qsgwl energy bands compared to Si.  (a) Si energy bands with green signify Si $p$
		character, red Si \emph{s} character.  (b) BAs bands with green signifying As $p$, red As $s$, and blue B
		\emph{s}+\emph{p} characters.  (c)
		The orbitally projected density of states in BAs.}
	\label{fig:electron}
\end{figure*}

\subsubsection{Electron and hole masses}A main focus of this work are the electron and hole masses.  We compute the electron masses at the conduction band
bottom near X, on the $\Gamma$-X line see Fig.~\ref{fig:electron} , and the Luttinger parameters calculated from
masses at the valence band maximum as

\begin{align}
	\gamma_{1}&={\frac{1}{2m_\mathrm{lh}^{001}}}+{\frac{1}{2m_\mathrm{hh}^{001}}} \nonumber \\
	\gamma_{2}&={\frac{1}{4m_\mathrm{lh}^{001}}}-{\frac{1}{4m_\mathrm{hh}^{001}}} \\
	\gamma_{3}&={\frac{1}{4m_\mathrm{lh}^{001}}}+{\frac{1}{4m_\mathrm{hh}^{001}}}-{\frac{1}{2m_\mathrm{hh}^{111}}} \nonumber
\end{align}

Our BSE electron masses are very similar to those of Buckeridge and Scanlon~\cite{buckeridge}. Also, it is notable that
higher levels of the theory only weakly modify the electron masses (see Table~\ref{tab:electron}).  Also shown is the
nonparabolicity parameter, defined as $m(E) = m^* (1 - \alpha E)$ (energy-dependent mass, with $m^*$ the mass at
$k{=}0$).

The split-off hole masses follow a pattern typical to zincblende semiconductors, with the former being essentially
isotropic ($m^*{\approx}0.24$) and ${m_\mathrm{hh}^{111}}{>}{m_\mathrm{hh}^{110}}{>}{m_\mathrm{hh}^{100}}$, but the
anisotropy is rather small (see $\gamma_3$, Table~\ref{tab:luttinger}).  Also notable is the smallness of $\gamma_2$, so
that along the 100 direction all three masses are nearly the same.  This is unusual, and is probably a consequence of the
large direct gap at $\Gamma$.  According to \qsgwl, the splitting of the  split-off band is 220\,meV, larger than Si (50\,meV) and
slightly smaller than smaller than AlAs (300\,meV) and GaAs (340\,meV).

The energy structure of BAs is quite similar to Si, both with the conduction band minimum on the $\Gamma$-X line near X.
Both Si and BAs pick up significant $d$ character near the conduction band minimum.  The masses in the two materials are
similar (Tables~\ref{tab:electron} and \ref{tab:luttinger}).  BAs is more polar than AlAs or GaAs: bands near the Fermi
level are heavily weighted to the anion in BAs: bands of B character are mostly far above $E_{F}$.

\begin{figure*}[htb]
	\includegraphics[scale=0.8]{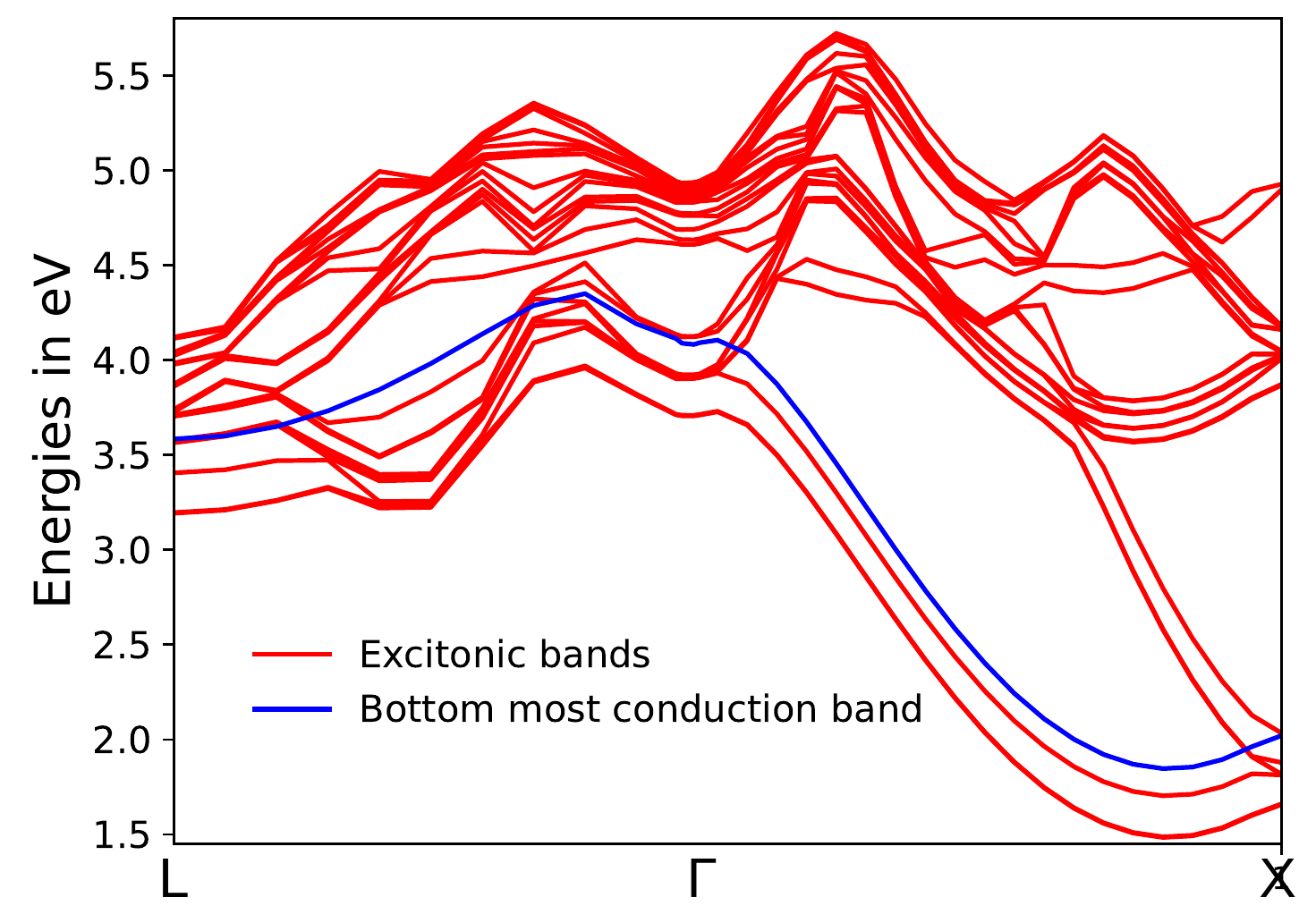}
	\caption{{\bf Excitonic band structure:} Exciton bands are shown along L$\Gamma$X path. The lowest conduction band is also shown. The deepest lying excitonic bands closely follow the conduction band profile. For excitons, the L, X and $\Gamma$ signify the length of the $\bf q$ vectors that connect the electrons to the holes and form excitons.}
	\label{exciton}
	
\end{figure*}

\subsection{Optical properties}We turn to the both macroscopic dielectric response and its \textbf{q}-dependence within the BSE. We observe bright
optical transitions at 4.0\,eV (Bai-Song et al.~\cite{song2020} reports 4.12 eV from their optical measurements),
however, we find dark optical transitions down to $\sim$1.5 eV. The deepest one is at the conduction band minimum along
the $\Gamma$-X path. These transitions are dipole forbidden and completely dark in a purely electronic framework. We
show the excitonic band structure in Fig.~\ref{exciton}. Symmetry lowering mechanisms, such as defects, spin-orbit
coupling, odd-parity phonon modes, Jan-Teller distortions can add to the oscillator strength of these otherwise dark
excitons. We also note the nontrivial role that spin-orbit coupling (SOC)
	plays in reducing the electronic and optical band gaps. We find that SOC reduces the electronic band gap by $\sim$90 meV
	and the optical gap by $\sim$220 meV. However, the oscillator strengths of the deepest lying dark exciton only gets weakly enhanced in presence of SOC. To put it in perspective, in a purely electronic framework, without considering the impact phonons and defects can have on these excitons, the oscillator strength of the 1.5 eV peak remains at least three orders of magnitude lower than the 1.2 eV exciton in CrI$_{3}$~\cite{acharya2022real,grzeszczyk2023strongly} and one order lower than the 1.6 eV exciton in NiO~\cite{acharya2022theory} and at least eight orders lower than the exciton in monolayer of MoS$_{2}$~\cite{acharya2022real}. Interestingly enough, Shin et al.~\cite{shin2022} observes a clear photo-luminescence peak at 695 nm (1.78 eV)
in BAs. This is consistent with our observation that there exists a rich spectrum of excitonic eigenvalues deep inside
the visible part of the optical spectrum, all the way down to the infrared edge. In particular, we observe a finite $\bf q$ ($\Gamma$-X) excitonic transition at 1.75 eV.  However, all such excitons are
essentially of Wannier-Mott character as the electrons and holes that for these excitons come from essentially the
conduction band bottom and valence band top~\cite{ref21}. We also could not find any mention of such excitonic peaks in the combined
experimental and theoretical optical studies from Bai-Song et al.~\cite{song2020}. We show in Fig.~\ref{exciton} the lowest conduction band in blue. The deepest lying excitonic band closely follows the conduction band profile. Along the entire path L$\Gamma$X the excitons remain bound by roughly $\sim$ 0.2-0.3 eV. Finally, within our approach we find
the dielectric constant $\epsilon_{\infty}$ to be 8.75, while in the theoretical literature we find numbers between 8 and
9 (see Table. 1 from Buckeridge and Scanlon~\cite{buckeridge}). We also note that as the e-h vertex screens the RPA W,
$\epsilon_{\infty}$ enhances from 6.55 to 8.75.

\section{Conclusions}Altogether, these observations suggest that our calculated one- and two-particle properties are in all likelihood the
correct intrinsic properties of cubic BAs.  Our observed electronic properties are in good agreement with experimental
observations wherever data from defect-free cubic BAs is available. Further, we predict light
hole masses and Luttinger parameters in BAs which might be significant considering the role BAs can play in semiconductor based
devising. Finally we observe a very rich series of dark excitonic lines deep inside the visible part of the optical
spectrum which can play a crucial role in application of cubic BAs in optoelectronics.

\medskip
\textbf{Supporting Information} \par 

We provide all input and output data files and commadlines to launch calculations that reproduce the results and figures from the paper at the deposition with digital object identifier~\cite{si}.  

\medskip
\textbf{Acknowledgements} \par 
MIK and SA are supported by the ERC Synergy Grant, project 854843 FASTCORR (Ultrafast dynamics of correlated electrons
in solids). MvS (and SA in the late stages of this work)  were
supported the by the Computational Chemical Sciences
program within the Office of Basic Energy Sciences,
U.S. Department of Energy under Contract No. DE-
AC36-08GO28308.  We acknowledge
PRACE for awarding us access to Irene-Rome hosted by TGCC, France and Juwels Booster and Cluster, Germany. Late stages
of calculations were performed using computational resources sponsored by the Department of Energy: the Eagle facility
at NREL, sponsored by the Office of Energy Efficiency and also the National Energy Research Scientific Computing Center,
under Contract No. DE-AC02-05CH11231 using NERSC award BES-ERCAP0021783.
\medskip

%

\begin{thebibliography}{29}%
	\makeatletter
	\providecommand \@ifxundefined [1]{%
		\@ifx{#1\undefined}
	}%
	\providecommand \@ifnum [1]{%
		\ifnum #1\expandafter \@firstoftwo
		\else \expandafter \@secondoftwo
		\fi
	}%
	\providecommand \@ifx [1]{%
		\ifx #1\expandafter \@firstoftwo
		\else \expandafter \@secondoftwo
		\fi
	}%
	\providecommand \natexlab [1]{#1}%
	\providecommand \enquote  [1]{``#1''}%
	\providecommand \bibnamefont  [1]{#1}%
	\providecommand \bibfnamefont [1]{#1}%
	\providecommand \citenamefont [1]{#1}%
	\providecommand \href@noop [0]{\@secondoftwo}%
	\providecommand \href [0]{\begingroup \@sanitize@url \@href}%
	\providecommand \@href[1]{\@@startlink{#1}\@@href}%
	\providecommand \@@href[1]{\endgroup#1\@@endlink}%
	\providecommand \@sanitize@url [0]{\catcode `\\12\catcode `\$12\catcode
		`\&12\catcode `\#12\catcode `\^12\catcode `\_12\catcode `\%12\relax}%
	\providecommand \@@startlink[1]{}%
	\providecommand \@@endlink[0]{}%
	\providecommand \url  [0]{\begingroup\@sanitize@url \@url }%
	\providecommand \@url [1]{\endgroup\@href {#1}{\urlprefix }}%
	\providecommand \urlprefix  [0]{URL }%
	\providecommand \Eprint [0]{\href }%
	\providecommand \doibase [0]{http://dx.doi.org/}%
	\providecommand \selectlanguage [0]{\@gobble}%
	\providecommand \bibinfo  [0]{\@secondoftwo}%
	\providecommand \bibfield  [0]{\@secondoftwo}%
	\providecommand \translation [1]{[#1]}%
	\providecommand \BibitemOpen [0]{}%
	\providecommand \bibitemStop [0]{}%
	\providecommand \bibitemNoStop [0]{.\EOS\space}%
	\providecommand \EOS [0]{\spacefactor3000\relax}%
	\providecommand \BibitemShut  [1]{\csname bibitem#1\endcsname}%
	\let\auto@bib@innerbib\@empty
	\bibitem [{\citenamefont {Shin}\ \emph {et~al.}(2022)\citenamefont {Shin},
		\citenamefont {Gamage}, \citenamefont {Ding}, \citenamefont {Chen},
		\citenamefont {Tian}, \citenamefont {Qian}, \citenamefont {Zhou},
		\citenamefont {Lee}, \citenamefont {Zhou}, \citenamefont {Shi} \emph
		{et~al.}}]{shin2022}%
	\BibitemOpen
	\bibfield  {author} {\bibinfo {author} {\bibfnamefont {J.}~\bibnamefont
			{Shin}}, \bibinfo {author} {\bibfnamefont {G.~A.}\ \bibnamefont {Gamage}},
		\bibinfo {author} {\bibfnamefont {Z.}~\bibnamefont {Ding}}, \bibinfo {author}
		{\bibfnamefont {K.}~\bibnamefont {Chen}}, \bibinfo {author} {\bibfnamefont
			{F.}~\bibnamefont {Tian}}, \bibinfo {author} {\bibfnamefont {X.}~\bibnamefont
			{Qian}}, \bibinfo {author} {\bibfnamefont {J.}~\bibnamefont {Zhou}}, \bibinfo
		{author} {\bibfnamefont {H.}~\bibnamefont {Lee}}, \bibinfo {author}
		{\bibfnamefont {J.}~\bibnamefont {Zhou}}, \bibinfo {author} {\bibfnamefont
			{L.}~\bibnamefont {Shi}},  \emph {et~al.},\ }\href@noop {} {\bibfield
		{journal} {\bibinfo  {journal} {Science}\ }\textbf {\bibinfo {volume}
			{377}},\ \bibinfo {pages} {437} (\bibinfo {year} {2022})}\BibitemShut
	{NoStop}%
	\bibitem [{\citenamefont {Lindsay}\ \emph {et~al.}(2013)\citenamefont
		{Lindsay}, \citenamefont {Broido},\ and\ \citenamefont {Reinecke}}]{theory2}%
	\BibitemOpen
	\bibfield  {author} {\bibinfo {author} {\bibfnamefont {L.}~\bibnamefont
			{Lindsay}}, \bibinfo {author} {\bibfnamefont {D.}~\bibnamefont {Broido}}, \
		and\ \bibinfo {author} {\bibfnamefont {T.}~\bibnamefont {Reinecke}},\
	}\href@noop {} {\bibfield  {journal} {\bibinfo  {journal} {Physical review
				letters}\ }\textbf {\bibinfo {volume} {111}},\ \bibinfo {pages} {025901}
		(\bibinfo {year} {2013})}\BibitemShut {NoStop}%
	\bibitem [{\citenamefont {Kang}\ \emph {et~al.}(2018)\citenamefont {Kang},
		\citenamefont {Li}, \citenamefont {Wu}, \citenamefont {Nguyen},\ and\
		\citenamefont {Hu}}]{tc1}%
	\BibitemOpen
	\bibfield  {author} {\bibinfo {author} {\bibfnamefont {J.~S.}\ \bibnamefont
			{Kang}}, \bibinfo {author} {\bibfnamefont {M.}~\bibnamefont {Li}}, \bibinfo
		{author} {\bibfnamefont {H.}~\bibnamefont {Wu}}, \bibinfo {author}
		{\bibfnamefont {H.}~\bibnamefont {Nguyen}}, \ and\ \bibinfo {author}
		{\bibfnamefont {Y.}~\bibnamefont {Hu}},\ }\href@noop {} {\bibfield  {journal}
		{\bibinfo  {journal} {Science}\ }\textbf {\bibinfo {volume} {361}},\ \bibinfo
		{pages} {575} (\bibinfo {year} {2018})}\BibitemShut {NoStop}%
	\bibitem [{\citenamefont {Li}\ \emph {et~al.}(2018)\citenamefont {Li},
		\citenamefont {Zheng}, \citenamefont {Lv}, \citenamefont {Liu}, \citenamefont
		{Wang}, \citenamefont {Huang}, \citenamefont {Cahill},\ and\ \citenamefont
		{Lv}}]{tc2}%
	\BibitemOpen
	\bibfield  {author} {\bibinfo {author} {\bibfnamefont {S.}~\bibnamefont
			{Li}}, \bibinfo {author} {\bibfnamefont {Q.}~\bibnamefont {Zheng}}, \bibinfo
		{author} {\bibfnamefont {Y.}~\bibnamefont {Lv}}, \bibinfo {author}
		{\bibfnamefont {X.}~\bibnamefont {Liu}}, \bibinfo {author} {\bibfnamefont
			{X.}~\bibnamefont {Wang}}, \bibinfo {author} {\bibfnamefont {P.~Y.}\
			\bibnamefont {Huang}}, \bibinfo {author} {\bibfnamefont {D.~G.}\ \bibnamefont
			{Cahill}}, \ and\ \bibinfo {author} {\bibfnamefont {B.}~\bibnamefont {Lv}},\
	}\href@noop {} {\bibfield  {journal} {\bibinfo  {journal} {Science}\ }\textbf
		{\bibinfo {volume} {361}},\ \bibinfo {pages} {579} (\bibinfo {year}
		{2018})}\BibitemShut {NoStop}%
	\bibitem [{\citenamefont {Tian}\ \emph {et~al.}(2018)\citenamefont {Tian},
		\citenamefont {Song}, \citenamefont {Chen}, \citenamefont {Ravichandran},
		\citenamefont {Lv}, \citenamefont {Chen}, \citenamefont {Sullivan},
		\citenamefont {Kim}, \citenamefont {Zhou}, \citenamefont {Liu} \emph
		{et~al.}}]{tc3}%
	\BibitemOpen
	\bibfield  {author} {\bibinfo {author} {\bibfnamefont {F.}~\bibnamefont
			{Tian}}, \bibinfo {author} {\bibfnamefont {B.}~\bibnamefont {Song}}, \bibinfo
		{author} {\bibfnamefont {X.}~\bibnamefont {Chen}}, \bibinfo {author}
		{\bibfnamefont {N.~K.}\ \bibnamefont {Ravichandran}}, \bibinfo {author}
		{\bibfnamefont {Y.}~\bibnamefont {Lv}}, \bibinfo {author} {\bibfnamefont
			{K.}~\bibnamefont {Chen}}, \bibinfo {author} {\bibfnamefont {S.}~\bibnamefont
			{Sullivan}}, \bibinfo {author} {\bibfnamefont {J.}~\bibnamefont {Kim}},
		\bibinfo {author} {\bibfnamefont {Y.}~\bibnamefont {Zhou}}, \bibinfo {author}
		{\bibfnamefont {T.-H.}\ \bibnamefont {Liu}},  \emph {et~al.},\ }\href@noop {}
	{\bibfield  {journal} {\bibinfo  {journal} {Science}\ }\textbf {\bibinfo
			{volume} {361}},\ \bibinfo {pages} {582} (\bibinfo {year}
		{2018})}\BibitemShut {NoStop}%
	\bibitem [{\citenamefont {Liu}\ \emph {et~al.}(2018)\citenamefont {Liu},
		\citenamefont {Song}, \citenamefont {Meroueh}, \citenamefont {Ding},
		\citenamefont {Song}, \citenamefont {Zhou}, \citenamefont {Li},\ and\
		\citenamefont {Chen}}]{theory1}%
	\BibitemOpen
	\bibfield  {author} {\bibinfo {author} {\bibfnamefont {T.-H.}\ \bibnamefont
			{Liu}}, \bibinfo {author} {\bibfnamefont {B.}~\bibnamefont {Song}}, \bibinfo
		{author} {\bibfnamefont {L.}~\bibnamefont {Meroueh}}, \bibinfo {author}
		{\bibfnamefont {Z.}~\bibnamefont {Ding}}, \bibinfo {author} {\bibfnamefont
			{Q.}~\bibnamefont {Song}}, \bibinfo {author} {\bibfnamefont {J.}~\bibnamefont
			{Zhou}}, \bibinfo {author} {\bibfnamefont {M.}~\bibnamefont {Li}}, \ and\
		\bibinfo {author} {\bibfnamefont {G.}~\bibnamefont {Chen}},\ }\href@noop {}
	{\bibfield  {journal} {\bibinfo  {journal} {Physical Review B}\ }\textbf
		{\bibinfo {volume} {98}},\ \bibinfo {pages} {081203} (\bibinfo {year}
		{2018})}\BibitemShut {NoStop}%
	\bibitem [{\citenamefont {Yue}\ \emph {et~al.}(2022)\citenamefont {Yue},
		\citenamefont {Tian}, \citenamefont {Sui}, \citenamefont {Mohebinia},
		\citenamefont {Wu}, \citenamefont {Tong}, \citenamefont {Wang}, \citenamefont
		{Wu}, \citenamefont {Zhang}, \citenamefont {Ren} \emph {et~al.}}]{ref11}%
	\BibitemOpen
	\bibfield  {author} {\bibinfo {author} {\bibfnamefont {S.}~\bibnamefont
			{Yue}}, \bibinfo {author} {\bibfnamefont {F.}~\bibnamefont {Tian}}, \bibinfo
		{author} {\bibfnamefont {X.}~\bibnamefont {Sui}}, \bibinfo {author}
		{\bibfnamefont {M.}~\bibnamefont {Mohebinia}}, \bibinfo {author}
		{\bibfnamefont {X.}~\bibnamefont {Wu}}, \bibinfo {author} {\bibfnamefont
			{T.}~\bibnamefont {Tong}}, \bibinfo {author} {\bibfnamefont {Z.}~\bibnamefont
			{Wang}}, \bibinfo {author} {\bibfnamefont {B.}~\bibnamefont {Wu}}, \bibinfo
		{author} {\bibfnamefont {Q.}~\bibnamefont {Zhang}}, \bibinfo {author}
		{\bibfnamefont {Z.}~\bibnamefont {Ren}},  \emph {et~al.},\ }\href@noop {}
	{\bibfield  {journal} {\bibinfo  {journal} {Science}\ }\textbf {\bibinfo
			{volume} {377}},\ \bibinfo {pages} {433} (\bibinfo {year}
		{2022})}\BibitemShut {NoStop}%
	\bibitem [{\citenamefont {Choudhry}\ \emph {et~al.}(2023)\citenamefont
		{Choudhry}, \citenamefont {Pan}, \citenamefont {He}, \citenamefont {Shaheen},
		\citenamefont {Kim}, \citenamefont {Gnabasik}, \citenamefont {Gamage},
		\citenamefont {Sun}, \citenamefont {Ackerman}, \citenamefont {Yang} \emph
		{et~al.}}]{ref12}%
	\BibitemOpen
	\bibfield  {author} {\bibinfo {author} {\bibfnamefont {U.}~\bibnamefont
			{Choudhry}}, \bibinfo {author} {\bibfnamefont {F.}~\bibnamefont {Pan}},
		\bibinfo {author} {\bibfnamefont {X.}~\bibnamefont {He}}, \bibinfo {author}
		{\bibfnamefont {B.}~\bibnamefont {Shaheen}}, \bibinfo {author} {\bibfnamefont
			{T.}~\bibnamefont {Kim}}, \bibinfo {author} {\bibfnamefont {R.}~\bibnamefont
			{Gnabasik}}, \bibinfo {author} {\bibfnamefont {G.~A.}\ \bibnamefont
			{Gamage}}, \bibinfo {author} {\bibfnamefont {H.}~\bibnamefont {Sun}},
		\bibinfo {author} {\bibfnamefont {A.}~\bibnamefont {Ackerman}}, \bibinfo
		{author} {\bibfnamefont {D.-S.}\ \bibnamefont {Yang}},  \emph {et~al.},\
	}\href@noop {} {\bibfield  {journal} {\bibinfo  {journal} {Matter}\ }\textbf
		{\bibinfo {volume} {6}},\ \bibinfo {pages} {206} (\bibinfo {year}
		{2023})}\BibitemShut {NoStop}%
	\bibitem [{\citenamefont {Buckeridge}\ and\ \citenamefont
		{Scanlon}(2019)}]{buckeridge}%
	\BibitemOpen
	\bibfield  {author} {\bibinfo {author} {\bibfnamefont {J.}~\bibnamefont
			{Buckeridge}}\ and\ \bibinfo {author} {\bibfnamefont {D.~O.}\ \bibnamefont
			{Scanlon}},\ }\href@noop {} {\bibfield  {journal} {\bibinfo  {journal}
			{Physical Review Materials}\ }\textbf {\bibinfo {volume} {3}},\ \bibinfo
		{pages} {051601} (\bibinfo {year} {2019})}\BibitemShut {NoStop}%
	\bibitem [{\citenamefont {Ku}(1966)}]{old1}%
	\BibitemOpen
	\bibfield  {author} {\bibinfo {author} {\bibfnamefont {S.}~\bibnamefont
			{Ku}},\ }\href@noop {} {\bibfield  {journal} {\bibinfo  {journal} {Journal of
				The Electrochemical Society}\ }\textbf {\bibinfo {volume} {113}},\ \bibinfo
		{pages} {813} (\bibinfo {year} {1966})}\BibitemShut {NoStop}%
	\bibitem [{\citenamefont {Chu}\ and\ \citenamefont {Hyslop}(1974)}]{old2}%
	\BibitemOpen
	\bibfield  {author} {\bibinfo {author} {\bibfnamefont {T.}~\bibnamefont
			{Chu}}\ and\ \bibinfo {author} {\bibfnamefont {A.}~\bibnamefont {Hyslop}},\
	}\href@noop {} {\bibfield  {journal} {\bibinfo  {journal} {Journal of The
				Electrochemical Society}\ }\textbf {\bibinfo {volume} {121}},\ \bibinfo
		{pages} {412} (\bibinfo {year} {1974})}\BibitemShut {NoStop}%
	\bibitem [{\citenamefont {Wang}\ \emph {et~al.}(2012)\citenamefont {Wang},
		\citenamefont {Swingle}, \citenamefont {Ye}, \citenamefont {Fan},
		\citenamefont {Cowley},\ and\ \citenamefont {Bard}}]{old3}%
	\BibitemOpen
	\bibfield  {author} {\bibinfo {author} {\bibfnamefont {S.}~\bibnamefont
			{Wang}}, \bibinfo {author} {\bibfnamefont {S.~F.}\ \bibnamefont {Swingle}},
		\bibinfo {author} {\bibfnamefont {H.}~\bibnamefont {Ye}}, \bibinfo {author}
		{\bibfnamefont {F.-R.~F.}\ \bibnamefont {Fan}}, \bibinfo {author}
		{\bibfnamefont {A.~H.}\ \bibnamefont {Cowley}}, \ and\ \bibinfo {author}
		{\bibfnamefont {A.~J.}\ \bibnamefont {Bard}},\ }\href@noop {} {\bibfield
		{journal} {\bibinfo  {journal} {Journal of the American Chemical Society}\
		}\textbf {\bibinfo {volume} {134}},\ \bibinfo {pages} {11056} (\bibinfo
		{year} {2012})}\BibitemShut {NoStop}%
	\bibitem [{\citenamefont {Kang}\ \emph {et~al.}(2019)\citenamefont {Kang},
		\citenamefont {Li}, \citenamefont {Wu}, \citenamefont {Nguyen},\ and\
		\citenamefont {Hu}}]{Kang19}%
	\BibitemOpen
	\bibfield  {author} {\bibinfo {author} {\bibfnamefont {J.~S.}\ \bibnamefont
			{Kang}}, \bibinfo {author} {\bibfnamefont {M.}~\bibnamefont {Li}}, \bibinfo
		{author} {\bibfnamefont {H.}~\bibnamefont {Wu}}, \bibinfo {author}
		{\bibfnamefont {H.}~\bibnamefont {Nguyen}}, \ and\ \bibinfo {author}
		{\bibfnamefont {Y.}~\bibnamefont {Hu}},\ }\href
	{https://aip.scitation.org/doi/10.1063/1.5116025} {\bibfield  {journal}
		{\bibinfo  {journal} {Applied Physics Letters.}\ }\textbf {\bibinfo {volume}
			{115}},\ \bibinfo {pages} {122103} (\bibinfo {year} {2019})}\BibitemShut
	{NoStop}%
	\bibitem [{\citenamefont {Song}\ \emph {et~al.}(2020)\citenamefont {Song},
		\citenamefont {Chen}, \citenamefont {Bushick}, \citenamefont {Mengle},
		\citenamefont {Tian}, \citenamefont {Gamage}, \citenamefont {Ren},
		\citenamefont {Kioupakis},\ and\ \citenamefont {Chen}}]{song2020}%
	\BibitemOpen
	\bibfield  {author} {\bibinfo {author} {\bibfnamefont {B.}~\bibnamefont
			{Song}}, \bibinfo {author} {\bibfnamefont {K.}~\bibnamefont {Chen}}, \bibinfo
		{author} {\bibfnamefont {K.}~\bibnamefont {Bushick}}, \bibinfo {author}
		{\bibfnamefont {K.~A.}\ \bibnamefont {Mengle}}, \bibinfo {author}
		{\bibfnamefont {F.}~\bibnamefont {Tian}}, \bibinfo {author} {\bibfnamefont
			{G.~A. G.~U.}\ \bibnamefont {Gamage}}, \bibinfo {author} {\bibfnamefont
			{Z.}~\bibnamefont {Ren}}, \bibinfo {author} {\bibfnamefont {E.}~\bibnamefont
			{Kioupakis}}, \ and\ \bibinfo {author} {\bibfnamefont {G.}~\bibnamefont
			{Chen}},\ }\href@noop {} {\bibfield  {journal} {\bibinfo  {journal} {Applied
				Physics Letters}\ }\textbf {\bibinfo {volume} {116}},\ \bibinfo {pages}
		{141903} (\bibinfo {year} {2020})}\BibitemShut {NoStop}%
	\bibitem [{\citenamefont {Mei}\ \emph {et~al.}(2022)\citenamefont {Mei},
		\citenamefont {Xia}, \citenamefont {Zhang}, \citenamefont {Wu}, \citenamefont
		{Chen}, \citenamefont {Ma}, \citenamefont {Kong}, \citenamefont {Peng},
		\citenamefont {Zhu},\ and\ \citenamefont {Zhang}}]{ref21}%
	\BibitemOpen
	\bibfield  {author} {\bibinfo {author} {\bibfnamefont {H.}~\bibnamefont
			{Mei}}, \bibinfo {author} {\bibfnamefont {Y.}~\bibnamefont {Xia}}, \bibinfo
		{author} {\bibfnamefont {Y.}~\bibnamefont {Zhang}}, \bibinfo {author}
		{\bibfnamefont {Y.}~\bibnamefont {Wu}}, \bibinfo {author} {\bibfnamefont
			{Y.}~\bibnamefont {Chen}}, \bibinfo {author} {\bibfnamefont {C.}~\bibnamefont
			{Ma}}, \bibinfo {author} {\bibfnamefont {M.}~\bibnamefont {Kong}}, \bibinfo
		{author} {\bibfnamefont {L.}~\bibnamefont {Peng}}, \bibinfo {author}
		{\bibfnamefont {H.}~\bibnamefont {Zhu}}, \ and\ \bibinfo {author}
		{\bibfnamefont {H.}~\bibnamefont {Zhang}},\ }\href@noop {} {\bibfield
		{journal} {\bibinfo  {journal} {Physical Chemistry Chemical Physics}\
		}\textbf {\bibinfo {volume} {24}},\ \bibinfo {pages} {9384} (\bibinfo {year}
		{2022})}\BibitemShut {NoStop}%
	\bibitem [{\citenamefont {van Schilfgaarde}\ \emph {et~al.}(2006)\citenamefont
		{van Schilfgaarde}, \citenamefont {Kotani},\ and\ \citenamefont
		{Faleev}}]{qsgw}%
	\BibitemOpen
	\bibfield  {author} {\bibinfo {author} {\bibfnamefont {M.}~\bibnamefont {van
				Schilfgaarde}}, \bibinfo {author} {\bibfnamefont {T.}~\bibnamefont {Kotani}},
		\ and\ \bibinfo {author} {\bibfnamefont {S.}~\bibnamefont {Faleev}},\
	}\href@noop {} {\bibfield  {journal} {\bibinfo  {journal} {Physical review
				letters}\ }\textbf {\bibinfo {volume} {96}},\ \bibinfo {pages} {226402}
		(\bibinfo {year} {2006})}\BibitemShut {NoStop}%
	\bibitem [{\citenamefont {Pashov}\ \emph {et~al.}(2020)\citenamefont {Pashov},
		\citenamefont {Acharya}, \citenamefont {Lambrecht}, \citenamefont {Jackson},
		\citenamefont {Belashchenko}, \citenamefont {Chantis}, \citenamefont
		{Jamet},\ and\ \citenamefont {van Schilfgaarde}}]{pashov2020questaal}%
	\BibitemOpen
	\bibfield  {author} {\bibinfo {author} {\bibfnamefont {D.}~\bibnamefont
			{Pashov}}, \bibinfo {author} {\bibfnamefont {S.}~\bibnamefont {Acharya}},
		\bibinfo {author} {\bibfnamefont {W.~R.}\ \bibnamefont {Lambrecht}}, \bibinfo
		{author} {\bibfnamefont {J.}~\bibnamefont {Jackson}}, \bibinfo {author}
		{\bibfnamefont {K.~D.}\ \bibnamefont {Belashchenko}}, \bibinfo {author}
		{\bibfnamefont {A.}~\bibnamefont {Chantis}}, \bibinfo {author} {\bibfnamefont
			{F.}~\bibnamefont {Jamet}}, \ and\ \bibinfo {author} {\bibfnamefont
			{M.}~\bibnamefont {van Schilfgaarde}},\ }\href
	{https://www.sciencedirect.com/science/article/pii/S0010465519303868}
	{\bibfield  {journal} {\bibinfo  {journal} {Computer Physics Communications}\
		}\textbf {\bibinfo {volume} {249}},\ \bibinfo {pages} {107065} (\bibinfo
		{year} {2020})}\BibitemShut {NoStop}%
	\bibitem [{\citenamefont {Ismail-Beigi}(2017)}]{variational}%
	\BibitemOpen
	\bibfield  {author} {\bibinfo {author} {\bibfnamefont {S.}~\bibnamefont
			{Ismail-Beigi}},\ }\href@noop {} {\bibfield  {journal} {\bibinfo  {journal}
			{Journal of Physics: Condensed Matter}\ }\textbf {\bibinfo {volume} {29}},\
		\bibinfo {pages} {385501} (\bibinfo {year} {2017})}\BibitemShut {NoStop}%
	\bibitem [{\citenamefont {Cunningham}\ \emph {et~al.}(2023)\citenamefont
		{Cunningham}, \citenamefont {Gruening}, \citenamefont {Pashov},\ and\
		\citenamefont {van Schilfgaarde}}]{Cunningham2023}%
	\BibitemOpen
	\bibfield  {author} {\bibinfo {author} {\bibfnamefont {B.}~\bibnamefont
			{Cunningham}}, \bibinfo {author} {\bibfnamefont {M.}~\bibnamefont
			{Gruening}}, \bibinfo {author} {\bibfnamefont {D.}~\bibnamefont {Pashov}}, \
		and\ \bibinfo {author} {\bibfnamefont {M.}~\bibnamefont {van Schilfgaarde}},\
	}\href {https://arxiv.org/abs/2302.06325} {\enquote {\bibinfo {title} {{QSGW:
					Quasiparticle Self consistent GW with ladder diagrams in W}},}\ } (\bibinfo
	{year} {2023}),\ \bibinfo {note} {preprint arXiv 2302.06325}\BibitemShut
	{NoStop}%
	\bibitem [{\citenamefont {Hirata}\ and\ \citenamefont
		{Head-Gordon}(1999)}]{hirata1999}%
	\BibitemOpen
	\bibfield  {author} {\bibinfo {author} {\bibfnamefont {S.}~\bibnamefont
			{Hirata}}\ and\ \bibinfo {author} {\bibfnamefont {M.}~\bibnamefont
			{Head-Gordon}},\ }\href
	{https://www.sciencedirect.com/science/article/pii/S0009261499011495}
	{\bibfield  {journal} {\bibinfo  {journal} {Chemical Physics Letters}\
		}\textbf {\bibinfo {volume} {314}},\ \bibinfo {pages} {291} (\bibinfo {year}
		{1999})}\BibitemShut {NoStop}%
	\bibitem [{\citenamefont {Acharya}\ \emph {et~al.}(2021)\citenamefont
		{Acharya}, \citenamefont {Pashov}, \citenamefont {Rudenko}, \citenamefont
		{R{\"o}sner}, \citenamefont {van Schilfgaarde},\ and\ \citenamefont
		{Katsnelson}}]{acharya2021importance}%
	\BibitemOpen
	\bibfield  {author} {\bibinfo {author} {\bibfnamefont {S.}~\bibnamefont
			{Acharya}}, \bibinfo {author} {\bibfnamefont {D.}~\bibnamefont {Pashov}},
		\bibinfo {author} {\bibfnamefont {A.~N.}\ \bibnamefont {Rudenko}}, \bibinfo
		{author} {\bibfnamefont {M.}~\bibnamefont {R{\"o}sner}}, \bibinfo {author}
		{\bibfnamefont {M.}~\bibnamefont {van Schilfgaarde}}, \ and\ \bibinfo
		{author} {\bibfnamefont {M.~I.}\ \bibnamefont {Katsnelson}},\ }\href@noop {}
	{\bibfield  {journal} {\bibinfo  {journal} {npj Computational Materials}\
		}\textbf {\bibinfo {volume} {7}},\ \bibinfo {pages} {1} (\bibinfo {year}
		{2021})}\BibitemShut {NoStop}%
	\bibitem [{\citenamefont {Hensel}\ \emph {et~al.}(1965)\citenamefont {Hensel},
		\citenamefont {Hasegawa},\ and\ \citenamefont {Nakayama}}]{Hensel65}%
	\BibitemOpen
	\bibfield  {author} {\bibinfo {author} {\bibfnamefont {J.~C.}\ \bibnamefont
			{Hensel}}, \bibinfo {author} {\bibfnamefont {H.}~\bibnamefont {Hasegawa}}, \
		and\ \bibinfo {author} {\bibfnamefont {M.}~\bibnamefont {Nakayama}},\ }\href
	{https://journals.aps.org/pr/abstract/10.1103/PhysRev.138.A225} {\bibfield
		{journal} {\bibinfo  {journal} {Phys. Rev.}\ }\textbf {\bibinfo {volume}
			{138}},\ \bibinfo {pages} {225} (\bibinfo {year} {1965})},\ \bibinfo {note}
	{si conduction masses are 0.91 and 0.19}\BibitemShut {NoStop}%
	\bibitem [{\citenamefont {Lautenschlager}\ \emph {et~al.}(1987)\citenamefont
		{Lautenschlager}, \citenamefont {Garriga}, \citenamefont {Vina},\ and\
		\citenamefont {Cardona}}]{Lautenschlager87}%
	\BibitemOpen
	\bibfield  {author} {\bibinfo {author} {\bibfnamefont {P.}~\bibnamefont
			{Lautenschlager}}, \bibinfo {author} {\bibfnamefont {M.}~\bibnamefont
			{Garriga}}, \bibinfo {author} {\bibfnamefont {L.}~\bibnamefont {Vina}}, \
		and\ \bibinfo {author} {\bibfnamefont {M.}~\bibnamefont {Cardona}},\ }\href
	{https://journals.aps.org/prb/abstract/10.1103/PhysRevB.36.4821} {\bibfield
		{journal} {\bibinfo  {journal} {Phys. Rev. B}\ }\textbf {\bibinfo {volume}
			{36}},\ \bibinfo {pages} {4821} (\bibinfo {year} {1987})}\BibitemShut
	{NoStop}%
	\bibitem [{Lan()}]{LandoltBornsteinVol41A1b}%
	\BibitemOpen
	\href {\doibase 10.1007/b80447} {\enquote {\bibinfo {title} {{Semiconductors
					· Group IV Elements, IV-IV and III-V Compounds. Part b - Electronic,
					Transport, Optical and Other Properties}},}\ }\bibinfo {note} {Copyright 1998
		Springer-Verlag Berlin Heidelberg}\BibitemShut {NoStop}%
	\bibitem [{\citenamefont {Vurgaftman}\ \emph {et~al.}(2001)\citenamefont
		{Vurgaftman}, \citenamefont {Meyer},\ and\ \citenamefont
		{Ram-Mohan}}]{Vurgaftman01}%
	\BibitemOpen
	\bibfield  {author} {\bibinfo {author} {\bibfnamefont {I.}~\bibnamefont
			{Vurgaftman}}, \bibinfo {author} {\bibfnamefont {J.~R.}\ \bibnamefont
			{Meyer}}, \ and\ \bibinfo {author} {\bibfnamefont {L.~R.}\ \bibnamefont
			{Ram-Mohan}},\ }\href {https://doi.org/10.1063/1.1368156} {\bibfield
		{journal} {\bibinfo  {journal} {Journal of Applied Physics}\ }\textbf
		{\bibinfo {volume} {89}},\ \bibinfo {pages} {5815} (\bibinfo {year}
		{2001})}\BibitemShut {NoStop}%
	\bibitem [{\citenamefont {Acharya}\ \emph
		{et~al.}(2022{\natexlab{a}})\citenamefont {Acharya}, \citenamefont {Pashov},
		\citenamefont {Rudenko}, \citenamefont {R{\"o}sner}, \citenamefont
		{Schilfgaarde},\ and\ \citenamefont {Katsnelson}}]{acharya2022real}%
	\BibitemOpen
	\bibfield  {author} {\bibinfo {author} {\bibfnamefont {S.}~\bibnamefont
			{Acharya}}, \bibinfo {author} {\bibfnamefont {D.}~\bibnamefont {Pashov}},
		\bibinfo {author} {\bibfnamefont {A.~N.}\ \bibnamefont {Rudenko}}, \bibinfo
		{author} {\bibfnamefont {M.}~\bibnamefont {R{\"o}sner}}, \bibinfo {author}
		{\bibfnamefont {M.~v.}\ \bibnamefont {Schilfgaarde}}, \ and\ \bibinfo
		{author} {\bibfnamefont {M.~I.}\ \bibnamefont {Katsnelson}},\ }\href@noop {}
	{\bibfield  {journal} {\bibinfo  {journal} {npj 2D Materials and
				Applications}\ }\textbf {\bibinfo {volume} {6}},\ \bibinfo {pages} {33}
		(\bibinfo {year} {2022}{\natexlab{a}})}\BibitemShut {NoStop}%
	\bibitem [{\citenamefont {Grzeszczyk}\ \emph {et~al.}(2023)\citenamefont
		{Grzeszczyk}, \citenamefont {Acharya}, \citenamefont {Pashov}, \citenamefont
		{Chen}, \citenamefont {Vaklinova}, \citenamefont {van Schilfgaarde},
		\citenamefont {Watanabe}, \citenamefont {Taniguchi}, \citenamefont
		{Novoselov}, \citenamefont {Katsnelson} \emph
		{et~al.}}]{grzeszczyk2023strongly}%
	\BibitemOpen
	\bibfield  {author} {\bibinfo {author} {\bibfnamefont {M.}~\bibnamefont
			{Grzeszczyk}}, \bibinfo {author} {\bibfnamefont {S.}~\bibnamefont {Acharya}},
		\bibinfo {author} {\bibfnamefont {D.}~\bibnamefont {Pashov}}, \bibinfo
		{author} {\bibfnamefont {Z.}~\bibnamefont {Chen}}, \bibinfo {author}
		{\bibfnamefont {K.}~\bibnamefont {Vaklinova}}, \bibinfo {author}
		{\bibfnamefont {M.}~\bibnamefont {van Schilfgaarde}}, \bibinfo {author}
		{\bibfnamefont {K.}~\bibnamefont {Watanabe}}, \bibinfo {author}
		{\bibfnamefont {T.}~\bibnamefont {Taniguchi}}, \bibinfo {author}
		{\bibfnamefont {K.}~\bibnamefont {Novoselov}}, \bibinfo {author}
		{\bibfnamefont {M.}~\bibnamefont {Katsnelson}},  \emph {et~al.},\ }\href@noop
	{} {\bibfield  {journal} {\bibinfo  {journal} {Advanced Materials}\ ,\
			\bibinfo {pages} {2209513}} (\bibinfo {year} {2023})}\BibitemShut {NoStop}%
	\bibitem [{\citenamefont {Acharya}\ \emph
		{et~al.}(2022{\natexlab{b}})\citenamefont {Acharya}, \citenamefont {Weber},
		\citenamefont {Pashov}, \citenamefont {van Schilfgaarde}, \citenamefont
		{Lichtenstein},\ and\ \citenamefont {Katsnelson}}]{acharya2022theory}%
	\BibitemOpen
	\bibfield  {author} {\bibinfo {author} {\bibfnamefont {S.}~\bibnamefont
			{Acharya}}, \bibinfo {author} {\bibfnamefont {C.}~\bibnamefont {Weber}},
		\bibinfo {author} {\bibfnamefont {D.}~\bibnamefont {Pashov}}, \bibinfo
		{author} {\bibfnamefont {M.}~\bibnamefont {van Schilfgaarde}}, \bibinfo
		{author} {\bibfnamefont {A.~I.}\ \bibnamefont {Lichtenstein}}, \ and\
		\bibinfo {author} {\bibfnamefont {M.~I.}\ \bibnamefont {Katsnelson}},\
	}\href@noop {} {\bibfield  {journal} {\bibinfo  {journal} {arXiv preprint
				arXiv:2204.11081}\ } (\bibinfo {year} {2022}{\natexlab{b}})}\BibitemShut
	{NoStop}%
	\bibitem [{si()}]{si}%
	\BibitemOpen
	\href {https://dx.doi.org/10.5281/zenodo.7923977} {\enquote {\bibinfo {title}
			{Input and output data for reproducing the results/figures from the paper are
				kept at},}\ }\BibitemShut {NoStop}%
\end{thebibliography}

%

\end{document}